# RNA-based Phylogenetic Methods:
# Application to Mammalian Mitochondrial RNA Sequences.

Cendrine Hudelot[1], Vivek Gowri-Shankar[2], Howsun Jow[2], Magnus Rattray[2] and Paul G. Higgs[3]

[1] School of Biological Sciences, University of Manchester, Manchester M13 9PT, UK.
[2] Department of Computer Science, University of Manchester, Manchester M13 9PL, UK.
[3] Department of Physics, McMaster University, Hamilton, Ontario L8S 4M1, Canada.

Contact details

Paul Higgs

Department of Physics, McMaster University, Hamilton, Ontario L8S 4M1, Canada.

Tel 1 905 525 9140 ext 26870

Fax 1 905 546 1252

email: higgsp@mcmaster.ca

**Abstract**

The PHASE software package allows phylogenetic tree construction with a number of evolutionary models designed specifically for use with RNA sequences that have conserved secondary structure. Evolution in the paired regions of RNAs occurs via compensatory substitutions, hence changes on either side of a pair are correlated. Accounting for this correlation is important for phylogenetic inference because it affects the likelihood calculation. In the present study we use the complete set of tRNA and rRNA sequences from 69 complete mammalian mitochondrial genomes. The likelihood calculation uses two evolutionary models simultaneously for different parts of the sequence: a paired-site model for the paired sites and a single-site model for the unpaired sites. We use Bayesian phylogenetic methods and a Markov chain Monte Carlo algorithm is used to obtain the most probable trees and posterior probabilities of clades. The results are well resolved for almost all the important branches on the mammalian tree. They support the arrangement of mammalian orders within the four supra-ordinal clades that have been identified by studies of much larger data sets mainly comprising nuclear genes. Groups such as the hedgehogs and the murid rodents, which have been problematic in previous studies with mitochondrial proteins, appear in their expected position with the other members of their order. Our choice of genes and evolutionary model appears to be more reliable and less subject to biases caused by variation in base composition than previous studies with mitochondrial genomes.

**Introduction**

Over the last decade, the use of molecular phylogenetic techniques has led to important changes in the way we understand the evolutionary tree of the mammals. The number of species for which appropriate sequence information is available has been increasing rapidly, and the methods being used have become increasingly sophisticated. This has led to a large interest in mammalian phylogenetics, as witnessed by the June 2002 symposium in Sorrento, and to an increasing degree of confidence in many features of the mammalian tree that were not appreciated only a few years ago.

One important feature that now appears to have strong support is that the mammalian orders can be divided into four principal supra-ordinal groups. This has been shown convincingly using large data sets mainly derived from nuclear genes (Madsen *et al.* 2001, Murphy *et al.* 2001a, 2001b). We will number these groups following Murphy *et al.* (2001a) as:

- group I: Afrotheria, containing Proboscidea, Sirenia, Hyracoidea, Tubulidentata, Macroscelidea and Afrosoricida;
- group II: Xenarthra;
- group III: referred to as either Supraprimates (Lin *et al.* 2002a) or Euarchontoglires (Murphy *et al.* 2001), containing Primates, Dermoptera, Scandentia, Rodentia, and Lagomorpha;
- group IV: usually referred to as Laurasiatheria (Waddell, Okada and Hasegawa 1999; Madsen *et al.* 2001), containing Cetartiodactyla, Perissodactyla, Carnivora, Pholidota, Chiroptera and Eulipotyphla.

It has taken some time for this picture to emerge. The identification of the Afrotheria group (Porter *et al.* 1996, Stanhope *et al.* 1998, Springer *et al.* 1999) was a surprise, since it contains orders which are morphologically diverse and superficially have little in common apart from their presumed African origin. The Xenarthra group (armadillos, sloths and anteaters) is South American and had been recognized as an early diverging group on morphological grounds. Due to the general difficulty of determining the root of the eutherian tree using molecular phylogenetics, the positioning of Xenarthra has been unstable, sometimes appearing quite deep within the eutherian radiation (Waddell *et al.* 1999; Cao *et al.* 2000). The present picture (Delsuc *et al.* 2002) puts Xenarthra back as an early branching group, but it is not possible to distinguish whether it branches before or after Afrotheria or as a sister group to Afrotheria. Supraprimates and Laurasiatheria are both diverse groups for which a lot of sequence data has been available for some time. These groups contain several species that have been a particular problem with molecular phylogenetics. The murid rodents (rats, mice and voles) and the hedgehogs have been repeatedly found to branch at the base of the eutherian tree in analysis of mitochondrial genomes (Penny *et al.* 1999; Arnason *et al.* 2002) and this would mean that the orders Rodentia and Eulipotyphla would not be monophyletic. However, the monophyly of these orders is supported by nuclear sequence data (Madsen *et al.* 2001, Murphy *et al.* 2001a, 2001b), and if the mitochondrial tree of the eutherians is left as unrooted (without addition of marsupials or monotremes) then there is much less apparent discrepancy between the mitochondrial and nuclear trees. It has therefore been argued that the positioning of the root in the mitochondrial trees is an artifact arising because some lineages contradict the assumption of a homogeneous and stationary substitution model (Lin *et al.* 2002a). Our own approach using RNA sequences from mitochondrial genomes appears to suffer less from these problems than most other methods (Jow *et al.* 2002).

It goes almost without saying that we are always interested in the 'hard' parts of the tree. These will be the nodes of the tree that are most ambiguous, and that are most sensitive to changes in methods and in evolutionary models. For this reason, we believe it is important to develop methods that can accurately describe the evolution of the particular types of sequence data that are being used. Our own interest centres on RNA sequences, such as rRNAs and tRNAs, that have conserved secondary structures. It is well known that compensatory substitutions occur frequently in the paired regions of RNA structures and that identification of compensatory changes in sequence alignments is one of the principal ways in which RNA secondary structures have been established (Gutell, 1996; Cannone *et al.* 2002). Several evolutionary models describing compensatory changes have been proposed (Schöniger & von Haeseler, 1994; Rzhetzky, 1995; Muse, 1995; Tillier & Collins, 1995,1998; Higgs 2000) that consider pairs of sites rather than single sites as the basic unit of evolution. We have recently carried out a comparison of these models using likelihood ratio tests (Savill *et al.* 2001).

A key feature of the compensatory substitution mechanism is that the rate of substitution depends on the thermodynamic stability of the intermediate state (Higgs, 2000; Savill *et al.* 2001). For example, GC pairs are observed to change more rapidly to AU pairs than CG pairs because in the first case the intermediate would be a GU pair, whereas in the second case the intermediate would be GG or CC, which would destabilize the helix. In some models used for RNA pairs, changes involving simultaneous double substitutions are disallowed on the grounds that the chance of simultaneous mutations on both sides of a pair is extremely small. However, it should be remembered that the evolutionary rate matrices describe the changes in the consensus sequence of the population, which is a result of mutation, selection, drift and fixation, and they do not simply represent the process of mutation in individual sequences. The population genetics theory of compensatory mutations (Kimura, 1985; Stephan, 1996; Higgs, 1998) explains how the double substitutions can arise via the one-step process described above. When sequence data is analysed (Savill *et al*. 2001, Higgs 2000) it is found that models that allow both single and double substitution rates fit the data much better than those that only allow single substitutions.

As far as phylogenetics is concerned, the most important point about compensatory substitutions is that they introduce correlations in the changes occurring in the two, paired sites. Almost all phylogenetic methods assume that substitutions at different sites are independent of one another. Strictly speaking, to use these methods with RNA sequences is invalid (but it is frequently done nevertheless). If sites are strongly correlated, but the correlation is ignored, then each piece of information is effectively counted twice. This leads to falsely high confidence levels in strongly supported clades in a tree, and falsely low support values for alternatives. Jow *et al*. (2002) give explicit examples of this. In view of the widespread use of RNA sequences (particularly rRNA) in phylogenetic studies in many groups of organisms, we felt it important to develop phylogenetic software that can deal more rigorously with correlations in paired sites. In our recent work (Jow *et al*. 2002), we introduced a program for Bayesian phylogenetic methods called PHASE ("PHylogenetics And Sequence Evolution"). This uses a new implementation of a Markov Chain Monte Carlo (MCMC) procedure to search tree space and allows a choice from a range of evolutionary models applicable to DNA and RNA sequence evolution. We applied it to a set of 54 Mammalian mitochondrial genomes, focussing on the paired regions of rRNA and tRNA genes. The results were obtained were very promising, with good resolution of some of the most important deep level nodes within the mammalian tree.

The method used in our previous paper had one important limitation. Although the first version of the PHASE software could deal with either single-site models or paired-site models, it could only do one of these at once. Therefore when using the paired regions of the sequence, it was necessary to eliminate all the unpaired sites. In the version of PHASE described here, we are able to use a paired-site and an unpaired-site model simultaneously in the likelihood calculation. Thus, we are able to make use of the whole of the RNA sequence. A second improvement in the present study is that several important additional mitochondrial genomes have become available. This study therefore includes representatives of the orders Sirenia, Macroscelidea, Pholidota and Dermoptera that were not in our previous study, plus additional species in some of the other orders.

**Materials and Methods**

*Sequence data*

The sequence data is taken from the complete set of tRNA and rRNA genes extracted from all mammalian genomes available in the NCBI database on August 6th, 2002. There are 69 taxa in total, including 62 placental mammals and an outgroup of 5 marsupials and 2 monotremes. The taxa and their NCBI accession numbers are listed in table 1. In particular, the following genomes have become available since our previous study (Jow *et al.* 2002): the echidna, *Tachyglossus aculeatus* and the wombat, *Vombatus ursinus* (Janke *et al.* 2002); the sea-lion, *Eumetopias jubatus*, the walrus, *Odobenus rosmarus*, the flying lemur, *Cynocephalus variegatus*, the hare, *Lepus europaeus*, the elephant shrew, *Macroscelides proboscideus*, the pangolin, *Manis tetradactyla*, the ring-tailed lemur, *Lemur catta*, the dugong, *Dugong dugon*, and the tamandua, *Tamandua tetradactyla* (Arnason *et al.* 2002); and three bear species *Ursus americanus, U. arctos,* and *U. maritimus* (Delisle & Strobeck, 2002). All the sequences and source references for the species used in this paper (together with many other metazoan species) can be obtained from our own database of mitochondrial genomes known as OGRe (Jameson *et al.* 2003), which also contains information on gene order in mitochondria.

We have used all 22 tRNAs and the small (12S) and large (16S) rRNAs. Missing tRNA Lys in three marsupials (the bandicoot, wallaroo and wombat) was treated as missing data in the analysis. The rRNA secondary structures were aligned by eye using the human rRNA structure from the Gutell lab as a guide (Cannone *et al.* 2002, http://www.rna.icmb.utexas.edu/). The mitochondrial tRNA profiles developed by Helm *et al.* (2000) were used as a guide to align the tRNAs. Most of the stem regions have highly conserved sequences, and the variations were carefully checked to allow, in at least 50% of taxa, a Watson-Crick or GU-UG pair. This criterion was chosen to conserve a large part of the molecules' structure and to allow enough flexibility for further addition of species at a later date. However, we also considered an alternative, stricter criterion in which stems were defined only at sites where 90% of species contained Watson-Crick or GU-UG pairs. This alternative criterion was used in order to measure the sensitivity of our results to the exact choice of cut-off used in defining our base-paired sites (see the Discussion section for details). Data not considered part of the stems under these criteria were included in the aligned loop regions. Loop regions that were very variable in length and could not be reliable aligned across all species were excluded from the analysis. The final data set for the 50% cutoff criterion contained 3571 nucleotides in total, consisting of 967 pairs and 1637 single sites. For the 90% cutoff, these numbers change to 862 pairs and 1847 single sites. The data set is available on request.

*Substitution models*

We use different substitution models for the stem and loop regions. For the loop regions we use the general time-reversible four-state model GTR4 (see, for example, Page and Holmes 1998, pp. 148-154). For the stem regions we use a model with seven states, six of which represent the most common base pairs (AU,GU,GC,UA,UG and CG), plus a composite mismatch state (MM) representing the other 10 less frequent pairs. We use the most general time-reversible seven-state model GTR7 (referred to as model 7A in Savill *et al.* 2001). Both models have been introduced previously and we do not go into detail here. The GTR4 model has 4 frequency parameters and 6 rate parameters defining the rate of substitution between distinct bases. However, there is a constraint that the frequencies must sum to one, which reduces the number of free parameters to 9. Similarly, the GTR7 model has 7 frequency parameters and 21 rate parameters, but the same constraint reduces the number of free parameters to 27.

Within both loop and stem regions we find significant variation in the substitution rate at different sites. We model this variation using the discrete-gamma model of Yang (1994). This model approximates the distribution of rates across different sites, using a number of discrete categories chosen to approximate a Gamma distribution. A single parameter determines the shape of this Gamma distribution. We use a different discrete-gamma model for the loops and stems respectively, with four rate categories in each case. This introduces two further parameters to be estimated.

*Bayesian phylogenetic inference*

The main features of our Bayesian inference algorithm have been described previously in Jow *et al.* (2002). We use an MCMC algorithm to sample from the space of topologies, branch lengths and substitution model parameters. The fraction of each topology appearing in the MCMC sample provides us with an estimate of the *posterior probability* of that topology given our model and given some prior probabilities for the parameters, branch lengths and topologies. We can also estimate the posterior probability distribution for the substitution model parameters and branch lengths, but we will typically summarise these quantities by their mean when reporting results. We use uniform priors on all quantities, as we have no strong prior preferences (we do not go into details here, see Jow *et al.* 2002). Two recent reviews provide a useful introduction to Bayesian phylogenetic inference using MCMC techniques (Huelsenbeck *et al.* 2001, Lewis 2001). The PHASE software and documentation are available from http://www.bioinf.man.ac.uk/resources/. The difference between the MCMC algorithm introduced by Jow *et al.* (2002) and the one used here is that we can now include an arbitrary number of different substitution models for different classes of site.

Rate models in phylogenetics are usually normalized to that there is one substitution per site per unit time. We normalize our GTR7 model for paired sites so that there is one substitution per pair per unit time (which can either be a double or single substitution). When the two models are used simultaneously, the rates in the GTR4 model for single sites are defined relative to the GTR7. Rates in the GTR4 model are first normalized within this model and are then multiplied by an additional parameter that determines the relative rate of evolution of single sites and pairs. We choose a uniform positive prior for this parameter with a cut-off at some upper limit to ensure the prior distribution is properly normalised. A Gaussian proposal distribution is used for this parameter with reflecting boundaries at zero and the upper limit defined by the prior.

Time in the MCMC simulations was measured in cycles. In each cycle we made one attempted change to a branch length and attempted to change one type of property of one of the two rate models. These property types are (i) all the frequency parameters, (ii) all the rate ratio parameters, (iii) the gamma shape parameter, or (iv) the relative rate of the GR4 to GR7 model. In addition, every 10 cycles one of the two topology-changing proposals (nearest neighbour interchange or subtree pruning and regrafting) was attempted with equal probability. Note that topology can also change as a result of the continuous branch length changes (see Jow *et al.* for a detailed description of these proposal mechanisms).

In order to check the consistency of the MCMC results we carried out four independent runs. For each run, the initial burn-in period was 1,400,000 cycles. The sampling period was then 400,000 cycles, during which a tree configuration was sampled every 10 cycles. This gives 40,000 trees from each run, and 160,000 trees in total. A long burn-in period was used because it was found that some parameters equilibriated much more slowly than others and we wanted to be confident that the simulations had reached equilibrium. With hindsight, we have improved the proposal mechanism so that the burn-in part of the simulation will be more efficient in future. The results from four independent MCMC runs were combined and summarised in the consensus tree shown in figure 1. The tree was constructed using the majority rule consensus methods implemented in PHYLIP (Felsenstein 1989) and numbers represent the posterior probability supporting each clade, *i.e.* the percentage of times that clade appears in the combined set of MCMC samples. Support values from all four runs were generally very consistent (see figure caption for details), even though they began from independent random starting configurations.

**Results**

Our results show good support for monophyly of the four supra-ordinal groups defined in the introduction. Groups I, II and III are found to be monophyletic with 100% posterior probability while group IV is monophyletic with 96% posterior probability. We find strong support for the sister relationship of groups III and IV (96%) but the relative positioning of I, II and (III,IV) is not well resolved. With 52% support we find groups I and II as sisters (as shown in figure 1), with 31% support we find group II branching first followed by group I and then (III,IV), and with 17% support we find group I branching first. These percentages have changed slightly from those in our previous analysis with only the paired regions (Jow *et al.* 2002), where the group I first alternative had less that 1% support. With our current results, we can therefore not rule out any of the three possibilities. It is interesting to note that we find weakest support for group I being the earliest branching group, while Murphy *et al.* (2001b) find 99% posterior support for this scenario in their Bayesian analysis. This question has also been addressed in detail by Delsuc *et al.* (2002), who also find the same three possibilities as us, but are unable to distinguish conclusively between them. The currently available sequence data thus seems to be insufficient to resolve this question.

Within group I, Afrotheria, we find 100% support for the sister relationship of the dugong and the elephant, in accordance with morphological and molecular results which place them in Paenungulata (Lavergne, 1996). The positioning of the elephant shrew is not well resolved. With highest probability it is found as a sister group to the Paenungulata (49%) but it is also possibly a sister species to the aardvark and tenrec (36%) or at the root of Afrotheria (14%). Several other studies (Murphy *et al.* 2001b, Delsuc *et al.* 2002, Arnason *et al.* 2002) find the arrangement (aardvark,(elephant shrew,tenrec)), but this arrangement occurs very rarely in our results. Instead, we find strong support (99%) for the pairing of between the aardvark and tenrec, to the exclusion of the elephant shrew. Afrotheria is a diverse group and there are still rather few complete mitochondrial genomes available. Better taxon sampling is likely to help resolve these questions.

Within group III we find 100% support for the rodents forming a monophyletic group. This is usually observed with nuclear sequences (e.g. Delsuc *et al.* 2002, Murphy *et al.* 2001b), and was also found in an early study of mitochondrial RNA sequences that used a very restricted set of species (Frye and Hedges, 1995). However, many previous studies with mitochondrial genomes in which the rodents have not been monophyletic due to the positioning of the murids close to the root of the tree (e.g. D'Erchia, 1996, Penny *et al.* 1999, Corneli & Ward, 2000, Arnason *et al.* 2002). It is likely that the apparent paraphyly of rodents in these studies is an artefact due to variation in the mutational mechanism in this lineage (see the arguments of Philippe, 1997, and Lin *et al.* 2002a). This study is the first using the full set of currently available mitochondrial sequences in which the rodents appear as a monophyletic group. This suggests that our choice of sequences and evolutionary model is less subject to biases than most previous methods used on mitochondrial sequences. There is a considerable improvement in resolution in the present study in comparison with our results using only the paired regions of the RNAs (Jow *et al.* 2002). In that case, both the murids and the other rodents appeared together in group III, but they were paraphyletic.

If we accept that the rodents are indeed monophyletic, the branching order of the three principal rodent groups is an important question. Our results (figure 1) give the murids as earliest branching and the squirrel group as sister to the guinea pig group. Support for all the branches within the rodents is 96-100%. However, given the tendency of the murids to move to the base of the tree, we do not find this completely convincing. Our result is the same as that of Lin *et al.* (2002a) in the case where they constrain the rodents to be monophyletic. In contrast, Madsen *et al.* (2001), Murphy *et al.* (2001a,b) and Delsuc *et al.* (2002) all have the squirrel group as earliest branching, although usually with fairly low support. Recent results using a variety of nuclear genes and a range of additional species are still unresolved on several important questions of phylogeny within the rodents (Adkins *et al.* 2001, Huchon *et al.* 2002).

The Glires group (rodents plus lagomorphs) has appeared in several recent studies (Murphy *et al.* 2001a,b; Lin *et al.* 2002a). We also find a close relationship between rodents and lagomorphs, however the tree shrew is also included in this group in our results: there is 100% support for a sister group relationship between tree shrew and lagomorphs. This differs from studies using mainly nuclear genes (Murphy *et al.*, 2001b) where the tree shrew has been placed close to the primate but is congruent with other mitochondrial DNA phylogenies (Schmitz *et al.* 2000; Lin, Waddell and Penny, 2002) where it has been suggested that the tree shrew may be problematic due to unusual base composition. The higher primates (anthropoids) are also a strongly supported monophyletic group. We find strong support for the sister relationship of the loris and ring-tailed lemur but the position of the tarsier is unresolved and variable. The tarsier is another problematic species in mitochondrial phylogenies (Schmitz *et al.* 2002), although evidence from SINES suggests that it should be closest to the anthropoids (Schmitz *et al.* 2001). The primates are found to be monophyletic with posterior probability 59% and therefore appear as a monophyletic group in the consensus tree (figure 1) with the flying lemur (Dermoptera) as a sister group. However, there is also some support (41%) for the flying lemur being placed within the primates, in a clade with the loris and ring-tailed lemur, which is why monophyly of the primates is not supported with 100% posterior probability. Using mitochondrial amino acid sequences, Arnason *et al.* (2002) also placed the flying lemur within the primates, as did Murphy *et al.* (2001a) with nuclear proteins. Madsen (2001), in contrast, has the flying lemur outside the primates but within group III. The position of Dermoptera is therefore still an interesting unresolved issue.

Within group IV we find strong support that all orders are monophyletic. Group IV contains the hedgehog/moonrat group, which is one of the most problematic in mitochondrial phylogenies, since it has unusual nucleotide and amino acid composition (Waddell *et al.* 1999). Rooted trees using mitochondrial sequences continue to find the hedgehog at the root of Eutheria (Cao *et al.* 2000; Arnason *et al.* 2002). This is almost certainly an artifact, and one recent study excludes the hedgehog completely due to these problems (Lin *et al.* 2002a). In contrast, when nuclear sequences are used, the hedgehog group is better behaved, and appears within Eulipotyphla with the moles and shrews (Murphy *et al.* 2001a,b; Madsen *et al.* 2001). It is therefore striking that Eulipotyphla appears as a monophyletic group in figure 1. This is the first study with mitochondrial sequences in which this is the case. As with the murids, this indicates that the methods we use here with RNA sequences are less biased by problems of non-stationarity in the rate matrix than the methods usually used with mitochondrial protein sequences. We obtain 96% support for the Eulipotyphla clade, whilst the hedgehog/moonrat pair is sister group to the shrew with 94% support, thus leaving the mole at the base of Eulipotyphla. This is equivalent to the arrangement obtain with nuclear genes (Murphy *et al.* 2001b).

We obtain Eulipotyphla as the earliest branching order within group IV, in agreement with nuclear sequence studies. This is a notable improvement in resolution over our previous study with only paired regions (Jow *et al*. 2002) where we found the eulipotyphlan species all in group IV, but in a poorly-resolved paraphyletic arrangement with the bats. The new grouping Fereuungulata (= Cetartiodactyla + Perissodactyla + Carnivora + Pholidota), proposed by Waddell *et al*. (1999), occurs in our consensus tree, but with only 65% support. This is mainly due to the mobility of the pangolin and because the relationship between Carnivora, Perissodactyla and Cetartiodactyla is not well resolved. The studies of Murphy *et al.* (2001b) and Arnason *et al.* (2002) strongly support the Carnivora with the pangolin as sister to Perissodactyla. We only find 6% support for this scenario and we also find relatively low support for Perissodactyla and Cetartiodactyla as sisters (65%). The most likely alternative, with 27% support, is that Carnivora and Perissodactyla are sisters. It is hoped that a larger sampling of the taxa in this part of the tree will stabilise the relationships in future studies. Unfortunately, however, Pholidota, of which the pangolin is a member, has only seven living species, which all belong to the same genus, and therefore this is likely to remain as a long branch. It is notable that the branching order of species within Cetartiodactyla is now much better resolved than our previous study (Jow *et al*. 2002), and is in agreement with other studies of this group (Gatesy *et al*. 1999).

**Discussion**

Trees obtained from mitochondrial proteins using similar species (Arnason *et al.* 2002) and the rooted trees in Lin *et al.* (2002a) differ from ours and are less congruent with the consensus emerging from studies using nuclear genes. One reason for the difference is probably in our choice of genes. In a previous study we used only the stems of all mitochondrial RNA genes (Jow *et al.* 2002) and also found support for the same four primary supra-ordinal clades, although resolution of other details was not as good as in the present study. Here we have added several species, as well as improving the method to combine both loops and helices. It is interesting to compare the results obtained with both helices and loops with those obtained using the two sets of sites separately. When only helices are used we obtain a tree very similar to that shown in Jow *et al.* (2002). The major clades I to IV are all fully resolved but some orders are paraphyletic within these clades. Eulipotyphla and Chiroptera mix together in a paraphyletic arrangement at the base of group IV, whilst within group III, the murids are separate from the remaining rodents, and the loris, lemur and tarsier are separate from the remaining primates. When only the loops are used, groups III and IV are no longer monophyletic. This is because the hedgehog and moonrat move right to the base of the eutherian tree (as with mitochondrial proteins) and because the lagomorphs move to become the second earliest branching group within the eutherians. With the exception of Eulipotyphla, all the orders are monophyletic with the loops only (including primates and rodents, which were paraphyletic with stems only). However, some of the orders have very low support values and the degree of resolution for the inter-ordinal nodes is very low. It is clear therefore, that combining stems and loops gives a better resolved and more reliable tree than either of the two sets of sites alone. In general, the stems contain more useful information for resolving the inter-ordinal relationships that the loops, however, loops add significant extra information that resolves some of the problems listed above when only stems are used.

It is also noticeable that the loop regions are significantly more variable in base composition that the stems, presumably because structural constraints disallow large changes in the base content on stems. It has been observed (Foster *et al.* 1997; Schmitz *et al.* 2002; Saccone *et al.* 2002) that this nucleotide bias in the codons affects the amino acid composition, so that phylogenetic methods using amino acids, codons and DNA may all be adversely affected. Since non-stationarity in the substitution model has been implicated as one reason for anomalous results in phylogenetic inference (Mooers and Holmes 2000) this may be significant, and it is probably the reason why the hedgehog moves to the base of the tree for both proteins and RNA-loops. Further work investigating non-stationarity of substitution models and mutational bias in mitochondrial genomes will be of great interest.

We find that the mean evolutionary rate for the loops and stems is almost identical, with the GTR4 model having an evolutionary rate 0.96 times that of the GTR7 rate. This is perhaps surprising since one would usually expect DNA within the loops to be under weaker evolutionary constraints. Springer *et al.* (1995) observed significantly faster substitution rates in the loops of mammalian rRNAs, for example. However, it should be remembered that we have removed many of the most variable sites in the loops from the analysis because they cannot be reliably aligned. Furthermore, for the single sites the unit of branch length is the substitution per site, whereas for the paired site it is a state-change per pair (where the state change can be either a single or double transition). For these reasons, it is difficult to accurately compare rates in helices and loops. The variable rate gamma distribution parameter is also quite similar for both models ($\alpha = 0.4$ for GTR4 and $\alpha = 0.58$ for GTR7) indicating a comparable variation in mean substitution rates over sites. One possible reason for the better resolution of the stem model for the earliest branch points is the larger alphabet. Another possibility is suggested when we look at the substitution rates in greater detail. Table 3 shows the substitution rates for the GTR4 model of the loops. We see that the rate of transitions is roughly double the rate of transversions. Table 5 shows the substitution rates for the GTR7 model, which was applied to the stems. There is a characteristic pattern showing relatively high substitution rates between AU, GU and GC on one hand, and between UA, UG and CG on the other. Within these two groups mutations can cause substitutions in the consensus sequence without passing through a mismatch pair and that is why these substitutions are preferred. However, substitutions between these groups are much slower and therefore provide information on a longer timescale.

In tables 2 and 4 we show the estimated state frequency parameters of the GTR4 and GTR7 substitution models respectively, and we also show the empirical state frequencies in the data set. The estimates for GTR4 differ from the empirical values and this might be improved with a non-stationary substitution model which could model compositional changes between species. The results for the GTR7 model show quite good agreement between the estimated and empirical values, except for the mismatch (MM) state, which has an estimated value about three times higher than the empirical value. This effect was explained by Jow *et al.* (2002) and is due to the combination of a base-pair model with a variable rates model. The rapidly evolving sites show a higher proportion of MM states. These sites also contribute more to the estimate of the frequencies because they provide less correlated samples from the state distribution, and this in turn biases the MM frequency estimate towards a higher value than typically observed. It was thought that our liberal cut-off criterion, defining a paired site as one in which 50% of species with a Watson-Crick or GU-UG pair, might exacerbate this problem. We therefore experimented with a data set constructed using a much stricter criterion with sites only assumed to be paired if Watson-Crick or GU-UG pairs were observed in at least 90% of species. The resulting inferred topology distribution was not greatly affected and the consensus tree was exactly the same as the one in figure 1, with similar support values. However, the estimated MM frequency was reduced to about 5% while the empirical frequency was reduced to about 1.5%. The difference in the empirical and estimated frequency was therefore not improved by using the stricter cut-off criterion. It was suggested in Jow *et al.* (2002) that we should use a different substitution model for each rate category in order to rectify this effect and we are currently considering how best to implement this.

In cases where both Bayesian and maximum likelihood (ML) methods are carried out on the same data, it is often found that the Bayesian posterior probabilities are significantly higher than the corresponding support values obtained by bootstrapping the ML method. For example, Murphy *et al.* (2001b) and Wilcox *et al.* (2002) observed this phenomenon and argued that the support values obtained by bootstrapping are too conservative and that the Bayesian posterior probabilities provide a better estimate of the confidence that can be attached to the inference. On the other hand, Suzuki *et al.* (2002) considered a simulation where equal numbers of sites are generated on three different trees and the data are concatenated. Bayesian methods often found strong support for one of the alternative trees, which they interpreted as a false positive. They therefore argue that posterior probabilities give overcredibility to the inferred result. Clearly more theoretical work is necessary to understand how to interpret both the posterior and bootstrap probabilities.

In our own case, we are using linked mitochondrial genes, so the possibility of different genes having different trees seems remote. It could also be argued that if one seriously suspects that concatenated nuclear genes evolved on different trees then one should be analysing them separately rather than concatenating them. In our view, the principal source of error in phylogenetics occurs when an inappropriate model is used to describe the data, rather than due to the concatenation of genes that have different trees. If the model is incorrect, then high likelihoods can be assigned to incorrect trees. This is clearly a problem for Bayesian methods, but it is equally a problem for every other method: bootstrapping does not tell you if the model used was inadequate. In our own results, we obtain many nodes with very high posterior probabilities (all the unlabelled nodes in the figure have 100% support).

We have used evolutionary models that are as realistic as possible for the RNA genes that we use. The posterior probabilities are therefore our best interpretation of the data *according to this model*. We cannot rule out the possibility that the results would change if more sophisticated features were added to the model. Bayesian methods are in fact a very practical way to allow the use of complex models with many parameters. We also wish to point out that the occurrence of the nodes with 100% support gives us confidence that our MCMC simulation is reaching equilibrium, since these same strongly supported clades arise during the burn-in periods of simulations starting in different random configurations.

**Conclusion**

Our work adds to the support of some of the most important aspects of the mammalian tree. We find strong support for monophyly of all established orders of placental mammals, with the possible exception of Dermoptera/Primates as discussed above. We also find strong support for four principal supra-ordinal clades, and good resolution is obtained at the species level within many of the larger orders. Taxa such as the murids and hedgehogs that have posed particular problems in previous analyses using mitochondrial genomes appear in what we take to be their correct positions in our analysis. We find that the rRNA and tRNA genes from mitochondrial genomes are extremely informative at many levels of the tree. The issues that are not resolved here are by and large not conclusively resolved by other data sets either, including some much larger sets of nuclear genes. Indeed, it is notable that our results are found using a relatively small data set of about 3.5 kb of DNA, compared with typical mitochondrial data sets of about 7-10kb. By introducing new evolutionary models that are specific to RNA sequence evolution and by using state-of-the-art MCMC methods for tree search, we have helped to resolve some of the contradictions between previous results based on mitochondrial and nuclear sequences. Our software package is now freely available, and we expect that these methods will prove useful for many other groups of organisms in addition to mammals.

**Acknowledgements**

Cendrine Hudelot and Magnus Rattray were supported by the Biotechnology and Biological Sciences Research Council of the UK. Paul Higgs is supported by the Canada Research Chairs organization.

**Tables**

| Classification | Scientific name | Common name | NCBI accession |
|---|---|---|---|
| Prototheria | | | |
| Ornithorynchidae | *Ornithorhynchus anatinus* | platypus | NC 000891 |
| Tachyglossidae | *Tachyglossus aculeatus* | Australian echidna | NC_003321 |
| Theria | | | |
| Metatheria (Marsupials) | | | |
| Didelphimorphia | *Didelphis virginiana* | North-American opossum | NC_001610 |
| Diprotodontia | *Macropus robustus* | wallaroo | NC_001794 |
| Diprotodontia | *Trichosurus vulpecula* | silver-gray possum | NC_003039 |
| Diprotodontia | *Vombatus ursinus* | common wombat | NC_003322 |
| Peramelemorphia | *Isoodon macrourus* | nothern brown bandicoot | NC_002746 |
| Eutheria (placentals) | | | |
| Afrosoricida | *Echinops telfairi* | small madagascar hedgehog (tenrec) | NC_002631 |
| Carnivora | *Arctocephalus forsteri* | New-Zealand fur seal | NC_004023 |
| Carnivora | *Canis familiaris* | dog | NC_002008 |
| Carnivora | *Eumetopias jubatus* | Steller sea lion | NC_004030 |
| Carnivora | *Felis catus* | cat | NC_001700 |
| Carnivora | *Halichoerus grypus* | gray seal | NC_001602 |
| Carnivora | *Odobenus rosmarus* | walrus | NC_004029 |
| Carnivora | *Phoca vitulina* | harbor seal | NC_001325 |
| Carnivora | *Ursus americanus* | American black bear | NC_003426 |
| Carnivora | *Ursus arctos* | brown bear | NC_003427 |
| Carnivora | *Ursus maritimus* | polar bear | NC_003428 |
| Cetartiodactyla | *Balaenoptera musculus* | blue whale | NC_001601 |
| Cetartiodactyla | *Balaenoptera physalus* | finback whale | NC_001321 |
| Cetartiodactyla | *Bos taurus* | cow | NC_001567 |
| Cetartiodactyla | *Hippopotamus amphibius* | hippopotamus | NC_000889 |
| Cetartiodactyla | *Lama pacos* | alpaca | NC_002504 |
| Cetartiodactyla | *Ovis aries* | sheep | NC_001941 |
| Cetartiodactyla | *Physeter catodon* | sperm whale | NC_002503 |
| Cetartiodactyla | *Sus scrofa* | pig | NC_000845 |
| Chiroptera | *Artibeus jamaicencis* | fruit eating bat | NC_002009 |
| Chiroptera | *Chalinolobus tuberculatus* | New-Zealand long-tailed bat | NC_002626 |
| Chiroptera | *Pteropus dasymallus* | Ryukyu flying fox | NC_002612 |
| Chiroptera | *Pteropus scapulatus* | little red flying fox | NC_002619 |
| Dermoptera | *Cynocephalus variegatus* | Malayan flying lemur | NC_004031 |
| Eulipotyphla | *Echinosorex gymnura* | moon rat | NC_004031 |
| Eulipotyphla | *Erinaceus europaeus* | western European hedgehog | NC_002080 |
| Eulipotyphla | *Soriculus fumidus* | shrew | NC_003040 |
| Eulipotyphla | *Talpa europaea* | European mole | NC_002391 |
| Lagomorpha | *Lepus europaeus* | hare | NC_004028 |
| Lagomorpha | *Ochotona collaris* | pika | NC_003033 |
| Lagomorpha | *Oryctolagus cuniculus* | rabbit | NC_001913 |
| Macroscelidea | *Macroscelides proboscideus* | short-eared elephant shrew | NC_004026 |
| Perissodactyla | *Ceratotherium simum* | white rhinoceros | NC_001808 |
| Perissodactyla | *Equus asinus* | ass | NC_001788 |
| Perissodactyla | *Equus caballu* | horse | NC_001640 |
| Perissodactyla | *Rhinoceros unicornis* | greater Indian rhinoceros | NC_001779 |
| Pholidota | *Manis tetradactyla* | long-tailed pangolin | NC_004027 |
| Primates | *Cebus albifrons* | white-fronted capuchin | NC_002763 |
| Primates | *Gorilla gorilla* | gorilla | NC_001645 |
| Primates | *Homo sapiens* | human | NC_001807 |
| Primates | *Hylobates lar* | common gibbon | NC_002082 |
| Primates | *Lemur catta* | ring-tailed lemur | NC_004025 |
| Primates | *Macaca sylvanus* | barbary ape | NC_002764 |
| Primates | *Nycticebus coucang* | slow loris | NC_002765 |
| Primates | *Pan paniscus* | pygmy chimpanzee | NC_001644 |
| Primates | *Pan troglodydes* | chimpanzee | NC_001643 |
| Primates | *Papio hamadryas* | baboon | NC_001992 |
| Primates | *Pongo pygmaeus* | orangutan | NC_001646 |
| Primates | *Tarsius bancanus* | western tarsier | NC_002811 |
| Proboscidea | *Loxodonta africana* | African savannah elephant | NC_000934 |
| Rodentia | *Cavia porcellus* | guinea pig | NC_000884 |
| Rodentia | *Mus musculus* | house mouse | NC_001569 |
| Rodentia | *Myoxus glis* | fat dormouse | NC_001892 |
| Rodentia | *Rattus norvegicus* | Norwegian rat | NC_001665 |
| Rodentia | *Sciurus vulgaris* | Eurasian red squirrel | NC_002369 |
| Rodentia | *Thryonomys swinderianus* | greater cane rat | NC_002658 |
| Rodentia | *Volemys kikuchii* | vole | NC_003041 |
| Scandentia | *Tupaia belangeri* | northern tree shrew | NC_002521 |
| Sirenia | *Dugong dugon* | dugong | NC_003314 |
| Tubulidentata | *Orycteropus afer* | aardvark | NC_002078 |
| Xenarthra | *Dasypus novemcinctus* | nine-banded armadillo | NC_001821 |
| Xenarthra | *Tamandua tetradactyla* | southern tamandua | NC_004032 |

Table 1: List of taxa included in the analysis with scientific names, common names and NCBI accession numbers.

|   | Empirical frequencies | Estimated frequencies |
|---|---|---|
| A | 0.4407 | 0.4852 |
| C | 0.1721 | 0.1602 |
| G | 0.1387 | 0.1082 |
| T | 0.2585 | 0.2464 |

Table 2: Mean posterior estimates of frequency parameters from the GTR4 model compared to the empirical frequency of nucleotides in the RNA loops.

|   | A | C | G | T |
|---|---|---|---|---|
| A | - | 0.1569 | 0.2472 | 0.1747 |
| C | 0.4753 | - | 0.0348 | 1.1953 |
| G | 1.1086 | 0.0515 | - | 0.2038 |
| T | 0.3441 | 0.7772 | 0.0895 | - |

Table 3: Mean posterior estimate of transition rate parameters from the GTR4 model.

|   | Empirical frequencies | Estimated frequencies |
|---|---|---|
| AU | 0.2488 | 0.2430 |
| GU | 0.0343 | 0.0331 |
| GC | 0.2414 | 0.1806 |
| UA | 0.2287 | 0.2346 |
| UG | 0.0316 | 0.0305 |
| CG | 0.1804 | 0.1677 |
| MM | 0.0347 | 0.1105 |

Table 4: Mean posterior estimate of frequency parameters from the GTR7 model compared to the empirical frequency of base-pairs in the RNA stems.

|    | *AU*   | *GU*   | *GC*   | *UA*   | *UG*   | *CG*   | *MM*   |
|----|--------|--------|--------|--------|--------|--------|--------|
| AU | -      | 0.2051 | 0.3682 | 0.0147 | 0.0021 | 0.0006 | 0.2809 |
| GU | 1.5060 | -      | 1.0638 | 0.0064 | 0.0029 | 0.0095 | 0.1647 |
| GC | 0.4954 | 0.1950 | -      | 0.0009 | 0.0009 | 0.0007 | 0.1463 |
| UA | 0.0153 | 0.0009 | 0.0007 | -      | 0.1613 | 0.2774 | 0.2498 |
| UG | 0.0164 | 0.0031 | 0.0052 | 1.2410 | -      | 0.6989 | 0.2336 |
| CG | 0.0009 | 0.0019 | 0.0007 | 0.3880 | 0.1271 | -      | 0.1793 |
| MM | 0.6177 | 0.0493 | 0.2391 | 0.5303 | 0.0645 | 0.2721 | -      |

Table 5: Mean posterior estimate of transition rate parameters from the GTR7 model.

**Figure 1 Caption**

Phylogeny of Mammals obtained using both paired and unpaired regions of mitochondrial rRNA and tRNA genes. Bayesian posterior probabilities are calculated using the PHASE program. Percentages shown are averaged over four independent MCMC runs. Internal nodes without numbers are supported with 100% posterior probability in all four runs. Percentages marked with an asterix are the most variable with percentages ranging between 5% and 7.5% from the mean over the four runs. The remaining numbers vary by less than 4% over all runs.

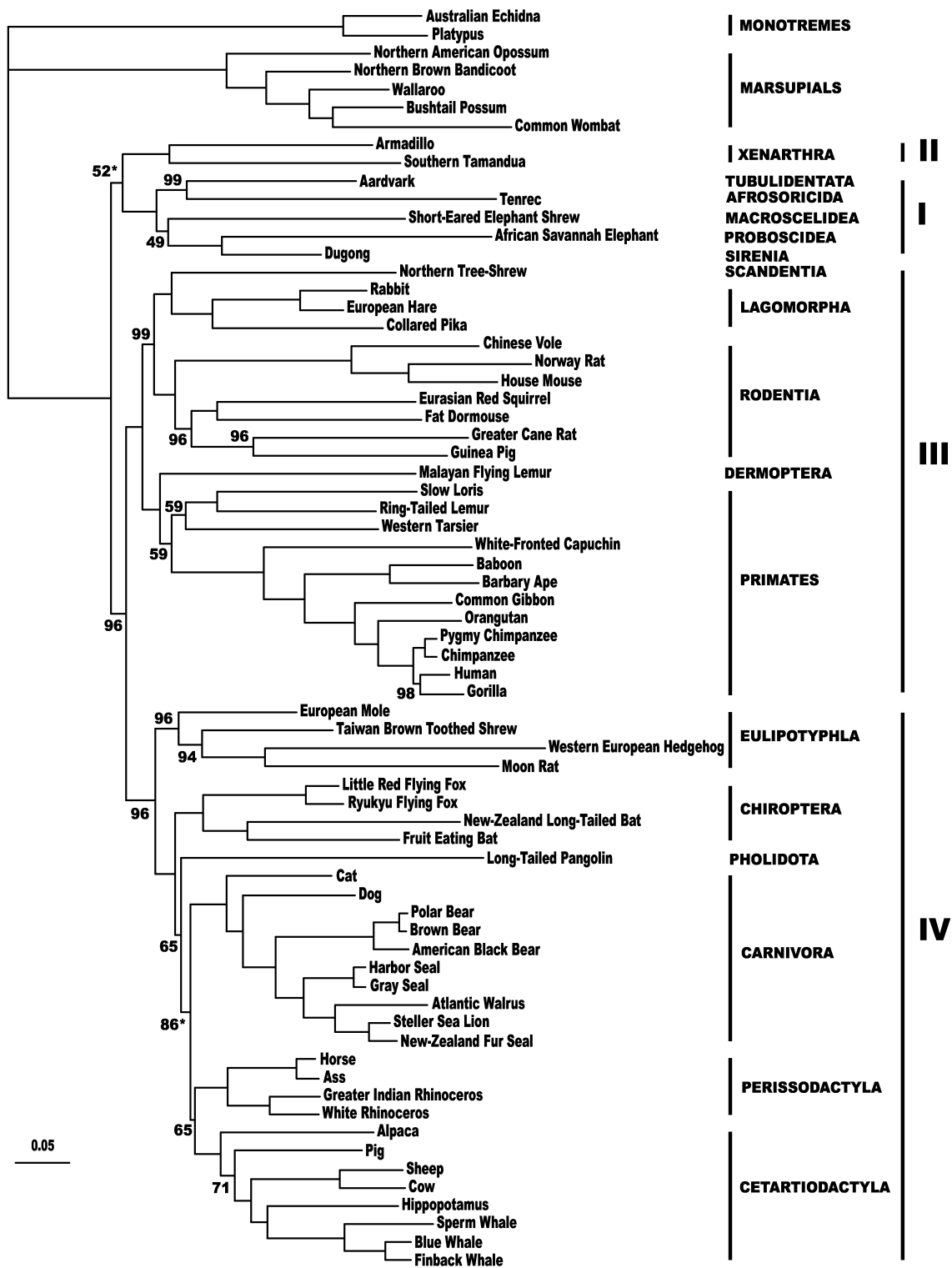